\documentclass[aps,prd,twocolumn,showpacs,superscriptaddress]{revtex4-2}
\usepackage{graphicx}
\usepackage{bm}
\usepackage{hyperref}
\usepackage{amsfonts}
\usepackage{mathptmx}
\usepackage{booktabs}
\usepackage{hhline}
\usepackage{amsmath}
\usepackage{xcolor}

\begin{document}

\title{A non-unitary solar constraint for long-baseline neutrino experiments}

\author{Andr\'es L\'opez Moreno}
\email{andres.lopezmoreno@lapp.in2p3.fr}
\affiliation{King's College London, Strand, London WC2R 2LS}
\affiliation{Laboratoire d'Annecy De Physique Des Particules, 9 Chem. de Bellevue, 74940 Annecy}

\begin{abstract}
Long-baseline neutrino oscillation experiments require external constraints on $\sin^2\theta_{12}$ and $\Delta m_{21}^2$ to make precision measurements of the leptonic mixing matrix. These constraints come from measurements of the Mikheyev-Smirnov-Wolfenstein (MSW) mixing in solar neutrinos. Here we develop an adiabatic MSW approximation in the presence of heavy neutral leptons; it adds, to leading order, a single new parameter ($\alpha_{11}$) representing the magnitude of the mixing between the $\nu_e$ state and the heavy sector. We use data from the Borexino, SNO and KamLAND collaborations to find a solar constraint appropriate for heavy neutral lepton searches in long-baseline oscillation experiments. Solar data yields a strongly correlated constraint on the solar mass splitting and the magnitude of $\nu_e$ non-unitary mixing, limiting the magnitude of the non-unitary parameter to $(1-\alpha_{11}) < 0.046$ at the $99\%$ credible region.
\end{abstract}


\maketitle

\section{\label{sec:intro} Introduction}

The next generation of long-baseline (LBL) neutrino experiments (DUNE\cite{DuneTDR2}, HK\cite{HKphysicsPotential}) will bring an era of unprecedented precision measurements to neutrino oscillation physics. This will enable further testing of the current 3-flavour paradigm~\cite{Giganti2017} against non-standard neutrino interaction theories and sterile neutrino hypotheses.

LBL experiments compare the $\nu_\mu (\bar{\nu}_\mu)$ and $\nu_e (\bar{\nu}_e)$ fluxes of a neutrino beam at a near and far detector and thus have, at most, 4 oscillation channels: $\nu_\mu(\bar{\nu}_\mu) \to \nu_\mu(\bar{\nu}_\mu)$, $\nu_e (\bar{\nu}_e) \to \nu_e (\bar{\nu}_e)$, $\nu_\mu \to \nu_e$ and $\bar\nu_\mu \to \bar\nu_e$. Having few channels presents a difficulty when trying to produce constraints on neutrino propagation models with a large number of parameters. Indeed, the Pontecorvo-Maki-Nakagawa-Sakata (PMNS) oscillation framework has 6 free parameters (two mass square differences, three mixing angles and one complex phase) and LBL experiments typically fix or impose external constraints on $\Delta m^2_{21}$ and $\theta_{12}$ to measure the remaining 4 parameters: $\theta_{13}, \theta_{23}, \Delta m^2_{32}$, and $\delta_{CP}$~\cite{novaResult}\cite{T2Knature}.

Other than agnostic probes of non-unitarity via goodness of fit comparisons~\cite{Ellis2020}, the search for neutrino oscillations with heavy neutral leptons (HNLs) is a natural target for LBL physics because it introduces few new free parameters. These HNLs are well-motivated by low-scale type-I seesaw models~\cite{lowscaleSeesaw} that give a satisfying explanation to the origin and smallness of neutrino masses ---allowing for deviations from unitarity by introducing mixing into new electroweak-scale leptons. Due to their large mass, we expect seesaw-scale HNLs not to be kinematically available, thus not taking part in oscillations. This should manifest as a deficit in the un-normalised neutrino flux for all flavours and at all baselines due to a portion of the active states immediately shifting into the sterile sector~\cite{Escrihuela2015}. In practice, near-detector normalisation hides the deficit in all but the appearance channels~\cite{Forero2021}.

The need for external solar constraints in long-baseline experiments becomes even greater when trying to set limits on non-unitary mixing models with additional degrees of freedom. In such setups, the usual solar constraint (which assumes unitarity) must be updated to be consistent with the non-unitary formalism. Currently, the lack of a non-unitary solar constraint is a hard wall for LBL analysers hoping to search for HNL mixing in oscillation data. This paper aims to produce one such constraint by finding a non-unitary expression for the adiabatic Mikheyev-Smirnov-Wolfenstein (MSW) effect in the presence of HNLs.

In section \ref{sec:solarConstraint} we discuss the relevance of the solar constraint on the PMNS parameters for LBL experiments. In section \ref{sec:nonUosc} we review the non-unitary neutrino mixing formalism and use it in deriving a non-unitary large mixing angle (LMA) MSW solution~\cite{Shi:1991zw}. Finally, in section \ref{sec:nonU-solar} we use the previous result to extract a new non-unitary solar constraint from Borexino~\cite{Borexino}, SNO~\cite{SNO}, and KamLAND~\cite{Piepke2001} data to be used in LBL non-unitary fits.

\section{\label{sec:solarConstraint} The solar constraint in LBL experiments}
\subsection{The need for a solar constraint}
LBL experiments can produce $\nu_\mu$ and $\bar\nu_\mu$ dominated beams by choosing the charge of decaying hadrons with a magnetic horn. These beams contain non-negligible $\nu_e$ and $\bar\nu_e$ fluxes which contribute to the oscillation signals in the far detector. In the 3 flavour ($3\nu$) PMNS model, the oscillations follow the familiar formula
\begin{eqnarray}
\begin{aligned}
    P_{\alpha \beta} = \delta_{\alpha \beta}\; -& \; 4\sum_{i>j}^3\mathbb{R}[U_{\alpha i} U^*_{\alpha j} U^*_{\beta i} U_{\beta j}]\sin^2{\frac{\Delta \hat{m}_{ij}^2L}{4E}} \\
    \; \pm & \; 2\sum_{i>j}^3\mathbb{I}[U_{\alpha i} U^*_{\alpha j} U^*_{\beta i} U_{\beta j}]\sin{\frac{\Delta \hat{m}_{ij}^2L}{2E}}
\end{aligned}
\label{eqn:vanillaOscillations}
\end{eqnarray}
where $(U)_{\alpha j}$ is a change of basis matrix between the flavour states and the propagation states, and $\hat{m}_{ij}^2$ are the eigenvalue square differences (in vacuum, these correspond to the entries of the PMNS matrix and the mass square differences $\Delta m_{ij}^2$). The sign of the final term differentiates between $\nu$ and $\bar{\nu}$.

At the first oscillation maximum, which is the target of LBL experiments, we require that the term governing the oscillation be close to 1. That is, we need $\sin^2(\Delta m_{32}^2L/4E) \approx \sin^2(\Delta m_{31}^2L/4E)\approx 1$. This yields $L/4E$ $\approx 630$ km$/$GeV; since $\Delta m_{21}^2 \sim \mathcal{O}(10^{-5})$ eV$^2$, we find that at this baseline $\sin^2(\Delta m_{21}^2L/4E)\sim\mathcal{O}(10^{-4})$. It follows that $\Delta m_{21}^2$ does not have a leading order contribution in disappearance channels around the first oscillation maximum and so, using  $\Delta m_{31}^2\approx \Delta m_{32}^2$ and disregarding terms $< \mathcal{O}(10^{-3}$) we arrive at the following expressions for long-baseline oscillation probabilities:
\begin{subequations}
  \begin{align}
    P^{LBL}_{\nu_e \to \nu_e} &\approx 1 - \sin^22\theta_{13}\sin^2\frac{\Delta m_{32}^2L}{4E} \\
    P^{LBL}_{\nu_\mu\to\nu_\mu} &\approx 1 - \sin^22\theta_{23}\sin^2\frac{\Delta m_{32}^2L}{4E} \\
    P^{LBL}_{\nu_\mu\to\nu_e} &\approx \sin^22\theta_{13}\sin^2\theta_{23}\sin^2\frac{\Delta m_{32}^2L}{4E} \nonumber \\
    & \;\;\;\;\;\; \pm \frac{\Delta m_{21}^2L}{4E} 8J_{CP} \sin^2\frac{\Delta m_{32}^2L}{4E}
  \end{align}
  \label{eqn:LBL_probs}
\end{subequations} \ \\
where $J_{CP}=s_{12}c_{12}s_{23}c_{23}s_{13}c_{13}^2\sin\delta_{CP}$  is the Jarlskog invariant. Here we see that the leading contribution of the solar parameters $\Delta m_{21}^2, \theta_{12}$ is in the appearance channels, and not separable from $\sin\delta_{CP}$. Indeed, LBL experiments have limited sensitivity to the solar parameters, and need an external constraint to make precision measurements of $\delta_{CP}$. A more detailed description of this effect and its consequences can be found in~\cite{Denton_2023}.

\begin{figure*}
\includegraphics[width=.47\textwidth]{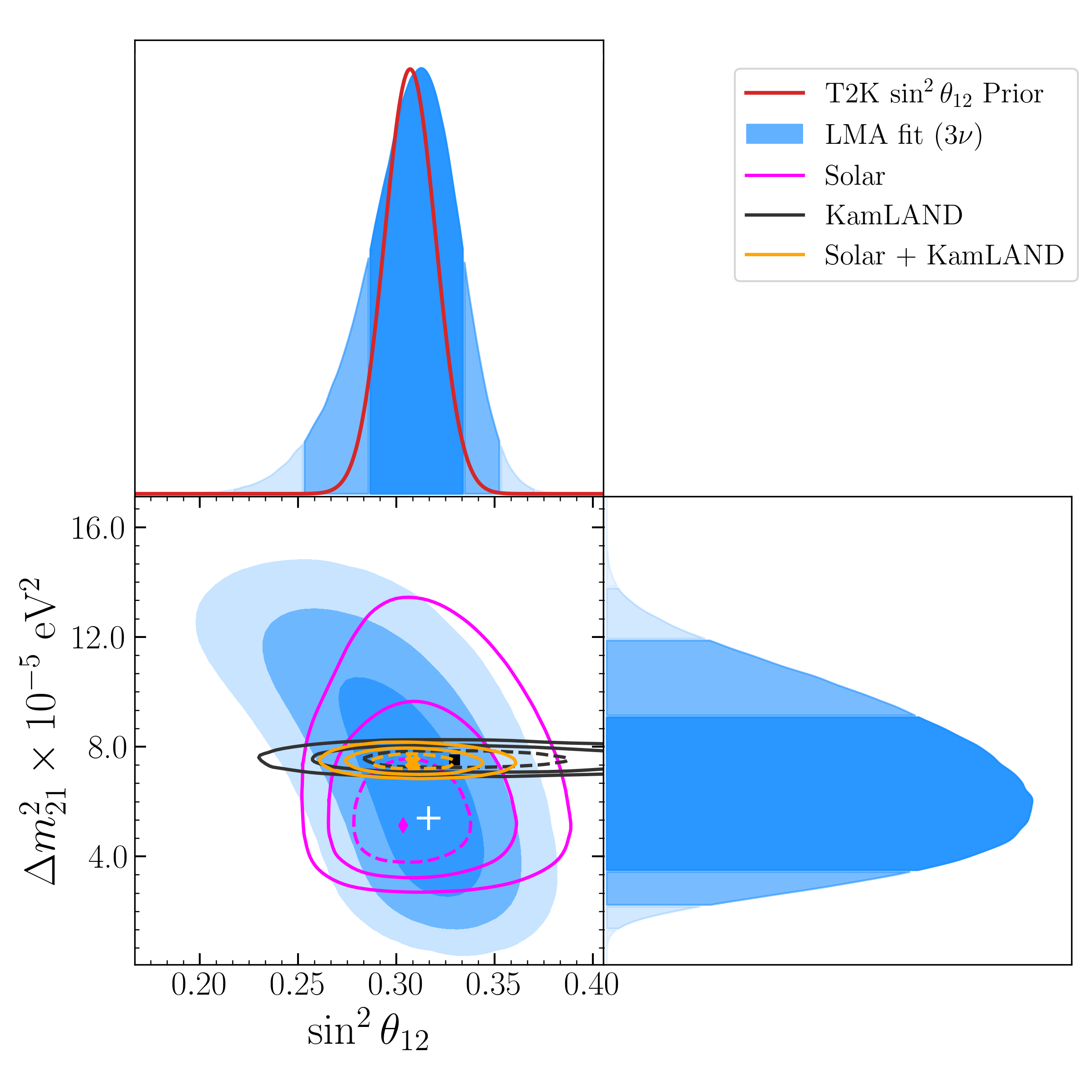}
\includegraphics[width=.47\textwidth]{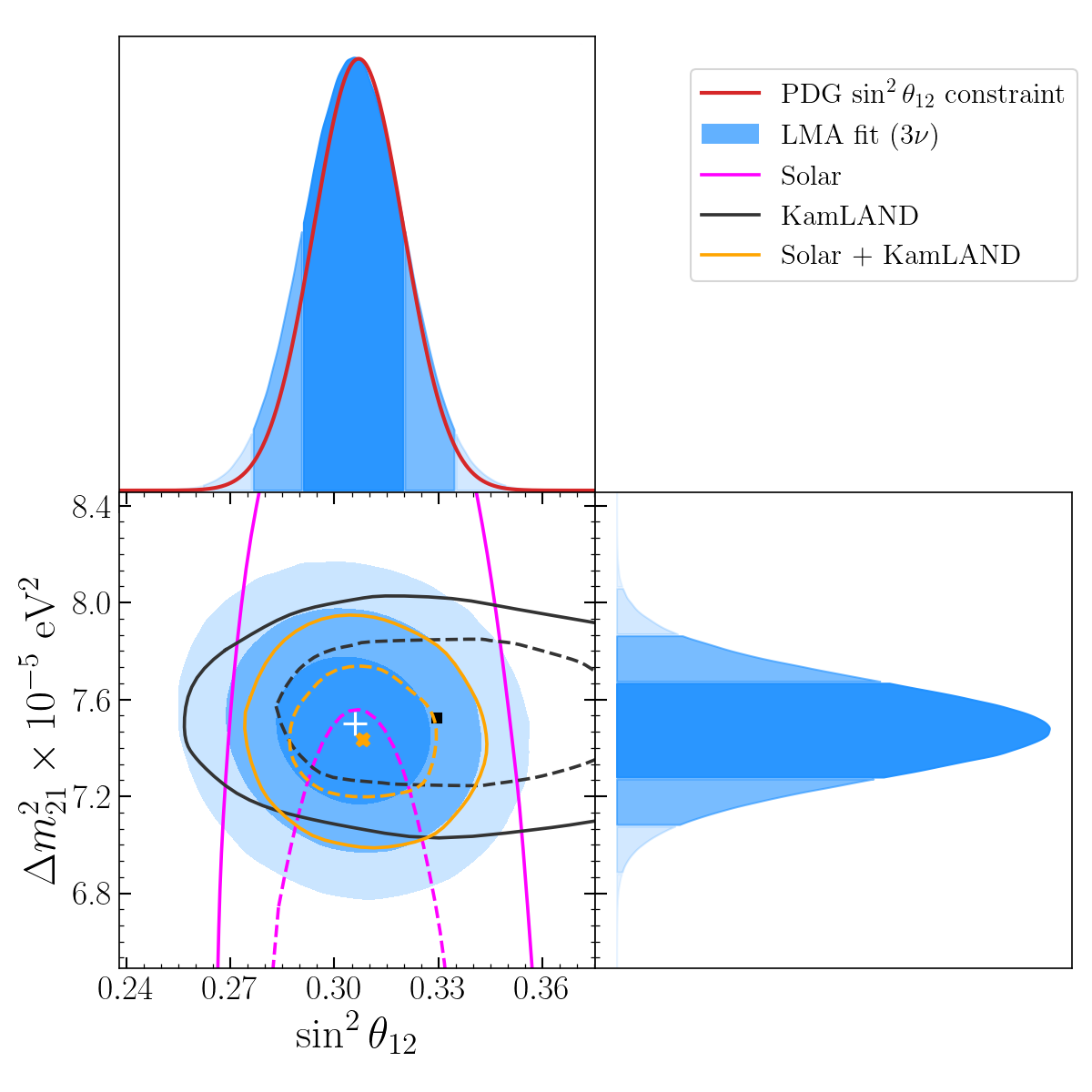}
\caption{\label{fig:3nu_fits} $1\sigma$, $2\sigma$ and $3\sigma$ contours of the posterior distributions for our Bayesian fit of the LMA solution with a uniform prior in both parameters (left) and a Gaussian prior on $ \Delta m_{21}^2 $ (right); with $1\sigma$, $2\sigma$ (right) and $3\sigma$ (left) contours for KamLAND's result (black), and the combined result from solar experiments (magenta). The red line corresponds to the $\sin^2\theta_{12}$ constraint recommended by the Particle Data Group and the highest posterior density point of our analysis is shown as a white cross.}
\end{figure*}

\subsection{Current bounds on solar neutrino parameters}
In the T2K~\cite{Abe2011} and NO$\nu$A~\cite{nova2021} experiments, the external constraint on the solar parameters comes from the 2019 Particle Data Group (PDG) report, which is derived from a combination of the KamLAND reactor data and various solar experiments~\cite{PDG2018}. KamLAND is a $\approx 1$ kton liquid scintillator detector which measures the $\bar{\nu_e}$ spectrum from 55 surrounding nuclear reactors which act as isotropic $\bar{\nu_e}$ sources in the $1-10$ MeV range. It provides constraints on $\Delta m_{21}^2$ and $\theta_{12}$ by fitting the 2-flavour $\bar{\nu_e}$ survival probability at a range of baselines where the frequency of the oscillations is dominated by $\Delta m_{21}^2$~\cite{Piepke2001}~\cite{Eguchi_2003}. To a very good approximation, the 3-flavour survival probability at KamLAND is given by
\begin{equation}
P(\bar{\nu}_e \to \bar{\nu}_e)\cong \cos^4\theta_{13}\Big[1-\sin^22\theta_{12}\sin^2\frac{\Delta m^2_{21}L}{4E_{\nu}} \Big]
\end{equation}
where $\theta_{13}$ is small and well-constrained. In this expression, $\theta_{12}$ gives the depth of the oscillation and $\Delta m^2_{21}$ the $L/E$ dependence; thus, KamLAND's $\Delta m_{21}^2$ and $\theta_{12}$ constraints are largely uncorrelated---this will prove helpful when redefining the angles in a non-unitary fit. Due to larger uncertainties in event rate predictions than in energy reconstruction~\cite{Abe2008}, KamLAND produces a strong constraint on the mass difference but a weaker constraint on the angle. In contrast, solar neutrino experiments produce comparatively weaker $\Delta m_{21}^2$ measurement but have better $\theta_{12}$ precision. In this paper, we consider solar neutrino data taken by the Borexino and SNO collaborations coming from p-p, pep, $^7$Be and $^8$B processes~\cite{Gann2021}.

Solar neutrino experiments measure $\nu_e$ survival probabilities by comparing predictions of the neutrino flux produced by nuclear processes in the sun~\cite{Asplund2009}\cite{Bahcall2004} with the measured flux on Earth. In the adiabatic regime ($\Delta m^2_{21} > 10^{-5} $eV$^2$), the survival probability can be approximated by a two-level transition between vacuum decoherent mixing and the MSW resonance~\cite{Kuo1986}. Explicitly, for the standard 3-flavour scenario,
\begin{eqnarray}
\begin{aligned}
    \cos2\hat\theta_{12} = &\frac{\cos2\theta_{12} - \beta}{\sqrt{(\cos2\theta_{12} - \beta)^2 + \sin^2(2\theta_{12})}} \\ 
    \\
    P_{\nu_e\to\nu_e} = \frac{1}{2}&\cos^4\theta_{13}[1 + \cos2\hat\theta_{12}\cos2\theta_{12}] + \sin^4\theta_{13}
\end{aligned}
    \label{eqn:LMA_3nu}
\end{eqnarray}
where $\hat\theta_{12}$ is the effective mixing angle near the transition point and $\beta$ is the ratio of matter to vacuum effect near the MSW eigenvalue crossing. $\beta$ can be written as
\begin{equation}
    \beta = \frac{2\sqrt{2}G_f\cos^2\theta_{13}n_e E}{\Delta m^2_{21}}
    \label{eqn:beta_3nu}
\end{equation}
for $n_e$ the electron density of the medium at production and $E$ the neutrino energy.

From equations \ref{eqn:LMA_3nu} and \ref{eqn:beta_3nu} we can see that a precise determination of $\theta_{12}$ can be made by measuring the survival probabilities in the limiting regimes of $\beta \to 0$ and $\beta > 1$, where $P_{\nu_e\to\nu_e}$ approaches 
\begin{equation}
P_{Vac} = \cos^4\theta_{13}[1 - \frac{1}{2}\sin^22\theta_{12}] + \sin^4\theta_{13}
\end{equation}
(vacuum averaged mixing), and 
\begin{equation}
P_{MSW} = \cos^4\theta_{13}\sin^2\theta_{12} + \sin^4\theta_{13}
\end{equation}
(MSW resonance) respectively.

This analysis uses publicly available data from solar neutrino experiments to fit the LMA curve to the reported survival probabilities for experiments at different energies. Survival probabilities for p-p, pep, $^7$Be and $^8$B from the Borexino collaboration are reported in~\cite{Agostini2018}, and a polynomial fit of the survival probability of $^8$B at energies near 10 MeV from the SNO collaboration is reported in~\cite{SNO-data}. While Borexino uses the high-metallicity Standard Solar Model (SSM)~\cite{Bahcall2005}, SNO uses the BS05 SSM~\cite{SNO-data}; these models show strong agreement in the neutrino production zones so the impact in the reported probabilities is small~\cite{SNO-thesis}. First, we show that this analysis is robust enough to recover the current solar constraints; this justifies using it to arrive at a sensible bound on non-unitary parameters.
\begin{figure}[h]
\centering
    \includegraphics[width=0.47\textwidth]{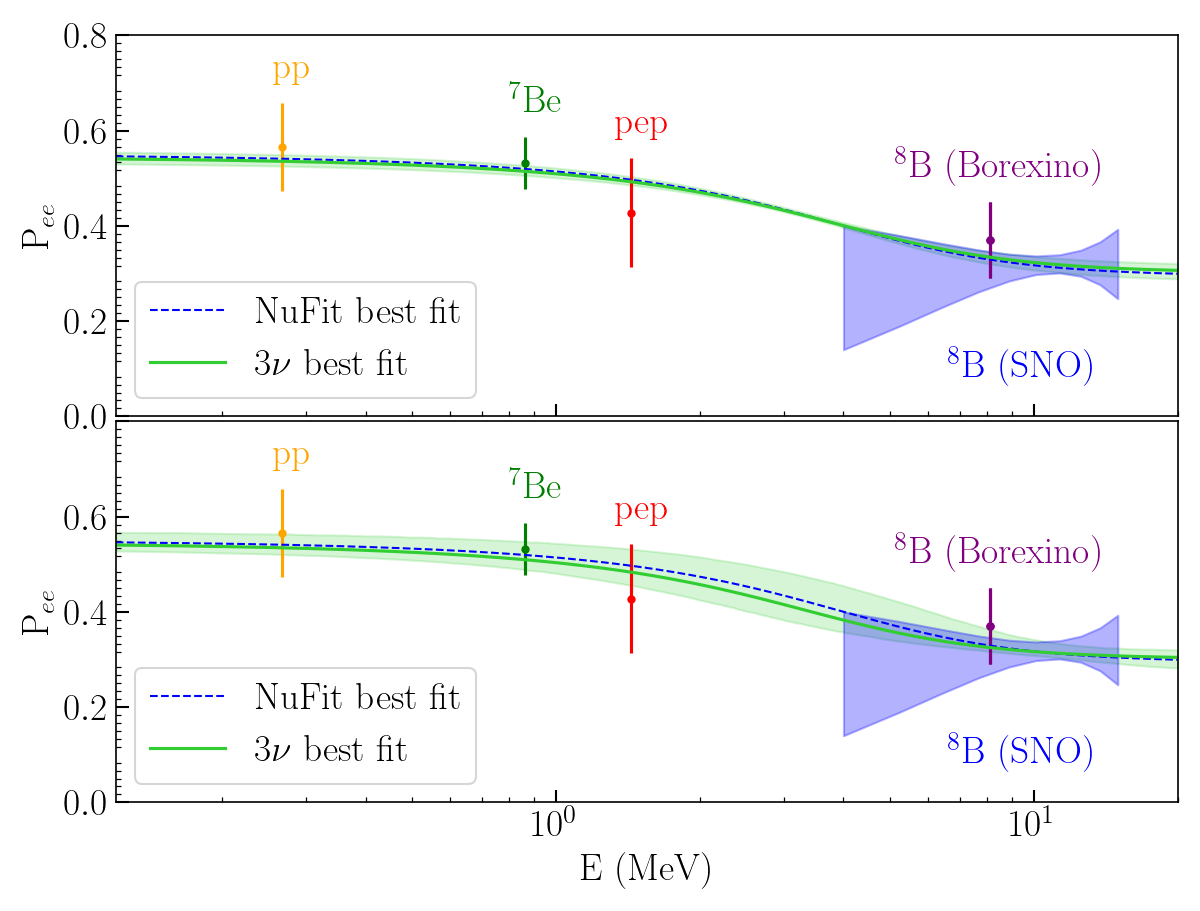}
    \caption{Solar survival probabilities from Borexino and SNO. The highest posterior density point and 1$\sigma$ contours for our Bayesian fit with (above) and without (below) a Gaussian prior on $\Delta m_{21}^2$ are shown in green. The LMA solution for NuFit's global best-fit values is shown in blue for reference.}
    \label{fig:3nu_survival}
\end{figure}

To this end, we perform two Bayesian fits of the survival probabilities described by equation \ref{eqn:LMA_3nu}; both with a Gaussian prior on $\sin^2\theta_{13}$ ($\mu = 0.022$, $\sigma=7 \times 10^{-4}$) coming from reactor experiments~\cite{NuFIT}\cite{DayaBay} and a uniform prior on $\sin^2\theta_{12}$. This is the prior recommended by the Particle Data Group~\cite{PDG2018}, and is consistent with the flavour-anarchic Haar prior on U(3)~\cite{EATON200287}. The first fit uses a uniform prior on $ \Delta m_{21}^2 $ and the second fit uses the KamLAND constraint as a Gaussian prior on $ \Delta m_{21}^2 $  ($\mu = 7.49 \times 10^{-5}$ eV$^2$, $\sigma=0.2 \times 10^{-5}$ eV$^2$)~\cite{kamland2011}. Both fits assume the same solar model used in SNO's analysis. Figure \ref{fig:3nu_fits} shows the posterior distributions of the fits with various overlaid constraints; the red curve corresponding to the PDG constraint is the prior used by the T2K experiment and is presented for reference. We find our fits to be in good agreement with the solar contours, but with weaker constraining power: without the KamLAND prior, the fit produces wider $\Delta m^2_{21}$ contours. This is not surprising because we have no access to Borexino's binned data, which would give a more precise constraint on the position of the MSW transition point. Despite this, the resulting distribution is consistent with the $\sin^2\theta_{12}$ constraint recommended by the Particle Data Group, particularly after applying the KamLAND prior.

Figure \ref{fig:3nu_survival} shows the fit results in terms of the $\nu_e$ survival probability, as well as the data used for the analysis.

\section{Non-unitarity in neutrino oscillations \label{sec:nonUosc}}
This analysis considers non-unitary effects due to charged-current transitions from the usual neutrinos to heavy isosinglet leptons (HNLs) which arise naturally from type-I seesaw models~\cite{Abdullahi2022}.  We adopt the formalism described by Escrihuela et al. in~\cite{Escrihuela2015}, where the heavy states separate from the oscillation at production and we observe a constant flux deficit in the active sector. Under this assumption, the effective $3\times 3$ mixing matrix between the active states is no longer unitary and can be parameterised by a lower-triangular matrix as
\begin{equation}
N = AU =
\begin{pmatrix}
    \alpha_{11} & 0 & 0 \\
    \alpha_{21} & \alpha_{22} & 0 \\
    \alpha_{31} & \alpha_{32} & \alpha_{33} 
\end{pmatrix} U
\label{eqn:nonU_mixingmatrix}
\end{equation}
where $U$ is the usual PMNS matrix\footnote{With (possibly) different mixing angles and CP-phase than in the unitary case.} and $A$ is the matrix encoding the non-unitary contribution to the oscillations. Only the off-diagonal elements of $A$ are allowed to have complex components. In this model, the matter potential has to be altered too so that, as shown in~\cite{Escrihuela2017}, the effective 3-flavour propagation Hamiltonian becomes
\begin{equation}
\mathcal{H}_{prop} = NMN^\dagger + (NN^\dagger)
\begin{pmatrix}
    v_{cc} - v_{nc} & 0 & 0 \\
    0 & -v_{nc} & 0 \\
    0 & 0 & -v_{nc} 
\end{pmatrix}
(NN^\dagger)
\label{eqn:nonU_hamiltonian}
\end{equation}
where $M$ is the matrix of mass eigenstates, $v_{cc} = \pm\sqrt{2}G_fN_e$ is the usual charged-current contribution and $v_{nc} = (\sqrt{2}/2)G_fN_n$ is the neutral-current contribution, which can no longer be ignored because mixing with the sterile states may give flavour sensitivity to the neutral-current potential. The choice of sign differentiates between $\nu$ and $\bar{\nu}$.
\subsection{Explicit form of the non-unitary propagation Hamiltonian}
 In the PDG parameterisation, the unitary lepton mixing matrix is given by $U=e^{i\theta_{23}\lambda_7}e^{i\theta_{13}\lambda_5}e^{i\theta_{12}\lambda_2}$ with a complex phase appended to the off-diagonal components of $e^{i\theta_{13}\lambda_5}$, where $\lambda_j$ are Gell-Mann matrices. Since we are only concerned with a survival probability $P(\nu_e \to \nu_e)$, we can use a similar argument to~\cite{Yokomakura_2002} to eliminate this complex phase $\delta_{CP}$.
 
 Inspired by the 3-flavour MSW derivation (\cite{Shi:1991zw}), we can further simplify the calculations by considering the propagation Hamiltonian in an alternate basis where we rotate by $e^{-i\theta_{23}\lambda_7}$ and $e^{-i\theta_{13}\lambda_5}$ so that the unitary part of the mixing matrix represents mixing between the mass states and some new non-flavour eigenstates. Usually, this reduces the unitary PMNS matrix to $U=e^{i\theta_{12}\lambda_2}$. Here, to leading order on the non-unitary corrections, we can commute the rotations with the lower-triangular matrix $A$ while only accruing small errors\footnote{Linear on the deviation from unitarity but suppressed by the smallness of $\theta_{13}$} on the electron row of the vacuum hamiltonian $\mathcal{H}_{vac(1i)}$. Thus,
 \begin{equation}
     \mathcal{H}_{vac} \approx Ae^{i\theta_{12}\lambda_2}Me^{-i\theta_{12}\lambda_2} A^\dagger
 \end{equation}

Similarly, the relevant entries of the matter contribution are largely invariant to commuting the non-unitary matrix $NN^\dag (= AA^\dag)$ with $e^{i\theta_{23}\lambda_7}$. This allows us to absorb the rotation into the unitary potential $V=- Diag(v_{nc}-v_{cc}, v_{nc}, v_{nc})$ for free, resulting in
 \begin{equation}
     \mathcal{H}_{mat} \approx e^{-i\theta_{13}\lambda_5}(AA^\dag) V (AA^\dag) e^{i\theta_{13}\lambda_5}
 \end{equation}

Factoring out the energy dependence, we find that the relevant entries of the traceless vacuum component become, to first order in the small parameters,
\begin{align}
    \mathcal{H}_{vac(11)} &\approx -\alpha_{11}^2\Delta m_{21}^2\cos2\theta_{12} \nonumber \\
    \mathcal{H}_{vac(12)} &\approx \alpha_{11}\Delta m_{21}^2(\alpha_{22}\sin2\theta_{12} - \overline{\alpha_{21}}\cos2\theta_{12}) \nonumber \\
    \mathcal{H}_{vac(13)} &\approx \alpha_{11}\Delta m_{21}^2(\overline{\alpha_{32}}\sin 2\theta_{12} - \overline{\alpha_{31}}\cos2\theta_{12}) \nonumber \\
    \mathcal{H}_{vac(22)} &\approx \alpha_{22}\Delta m_{21}^2(\alpha_{22}\cos2\theta_{12} + \mathcal{O}(|\alpha_{21}|))
\label{eqn:vacuum_hamiltonian_entries}
\end{align}
and, in the matter component,
\begin{align}
    \mathcal{H}_{mat(11)} &\approx \alpha_{11}^4c_{13}^2(v_{cc}-v_{nc}) \nonumber \\&\;\;\;\;- \alpha_{11}^3\sin2\theta_{13}\mathfrak{Re}\{\alpha_{31}\}(v_{cc}-2v_{nc}) \nonumber \\
    \mathcal{H}_{mat(12)} &\approx \alpha_{11}^3\Big( c_{13}\overline{\alpha_{21}}(v_{cc} - 2v_{nc}) - 2s_{13}\overline{\alpha_{32}}v_{nc}\Big) \nonumber \\ 
    \mathcal{H}_{mat(13)} &\approx s_{13}c_{13}(\alpha_{11}^4-\alpha_{33}^4)(v_{cc}-v_{nc}) \nonumber \\&\;\;\;\;+ c_{13}^2\alpha_{11}^3\overline{\alpha_{31}}(v_{cc}-2v_{nc}) \nonumber \\
    \mathcal{H}_{mat(22)} &\approx -\alpha_{22}^4v_{nc}.
\label{eqn:matter_hamiltonian_entries}
\end{align}
\begin{figure}
\centering
    \includegraphics[width=0.47
    \textwidth]{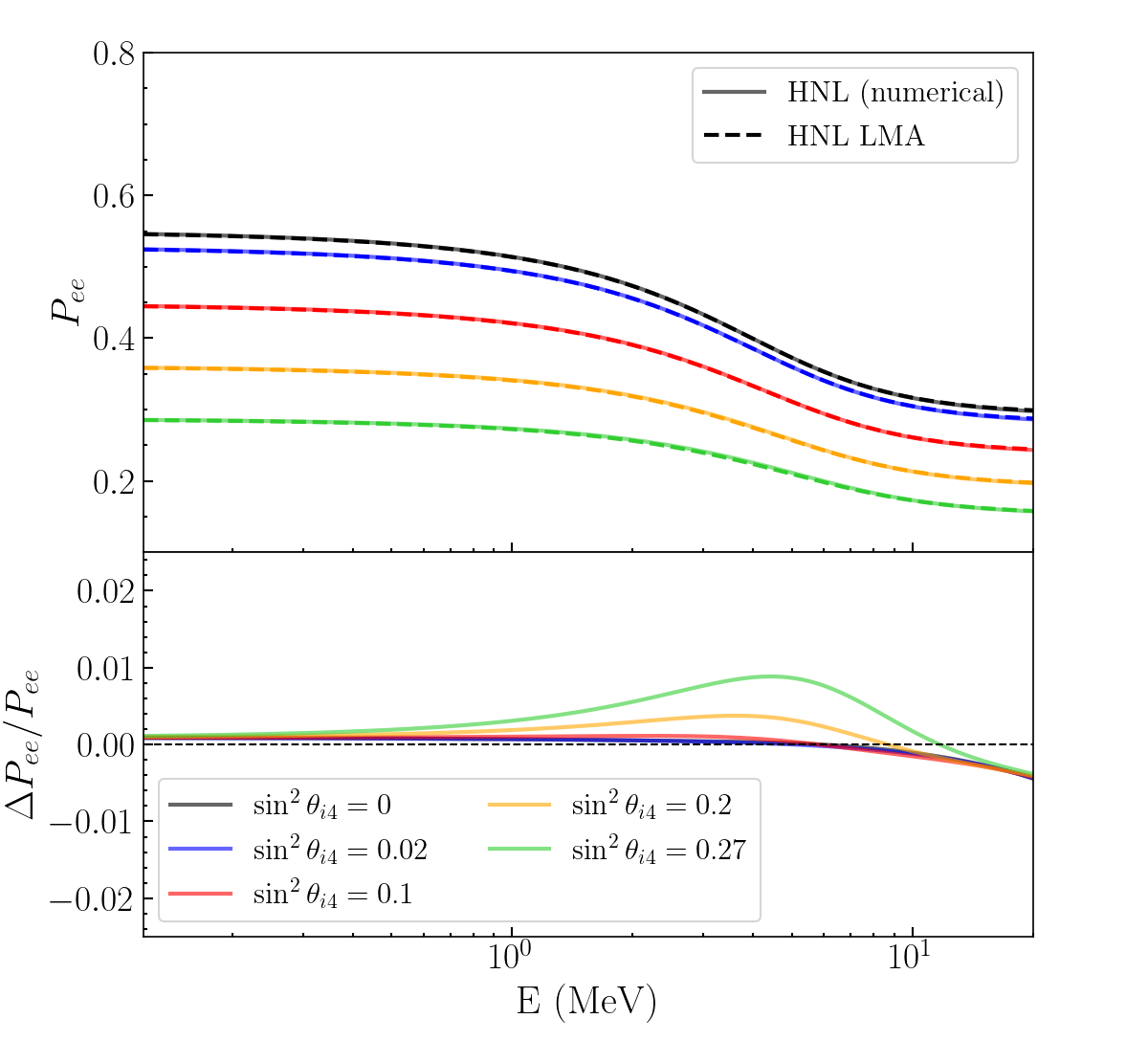}
    \caption{Comparison of the usual (black) and non-unitary (coloured) LMA approximations against numerical solutions in the presence of a single GeV scale HNL, assuming the electron and neutron densities of the sun's core. The magnitude of the mixing into the sterile sector was set to the same value for all three active neutrinos. In this case, the non unitary parameter $\alpha_{11}$ corresponds to $\cos\theta_{14}$.}
    \label{fig:LMAtest}
\end{figure}
\begin{figure*}[t!]
\includegraphics[width=.47\textwidth]{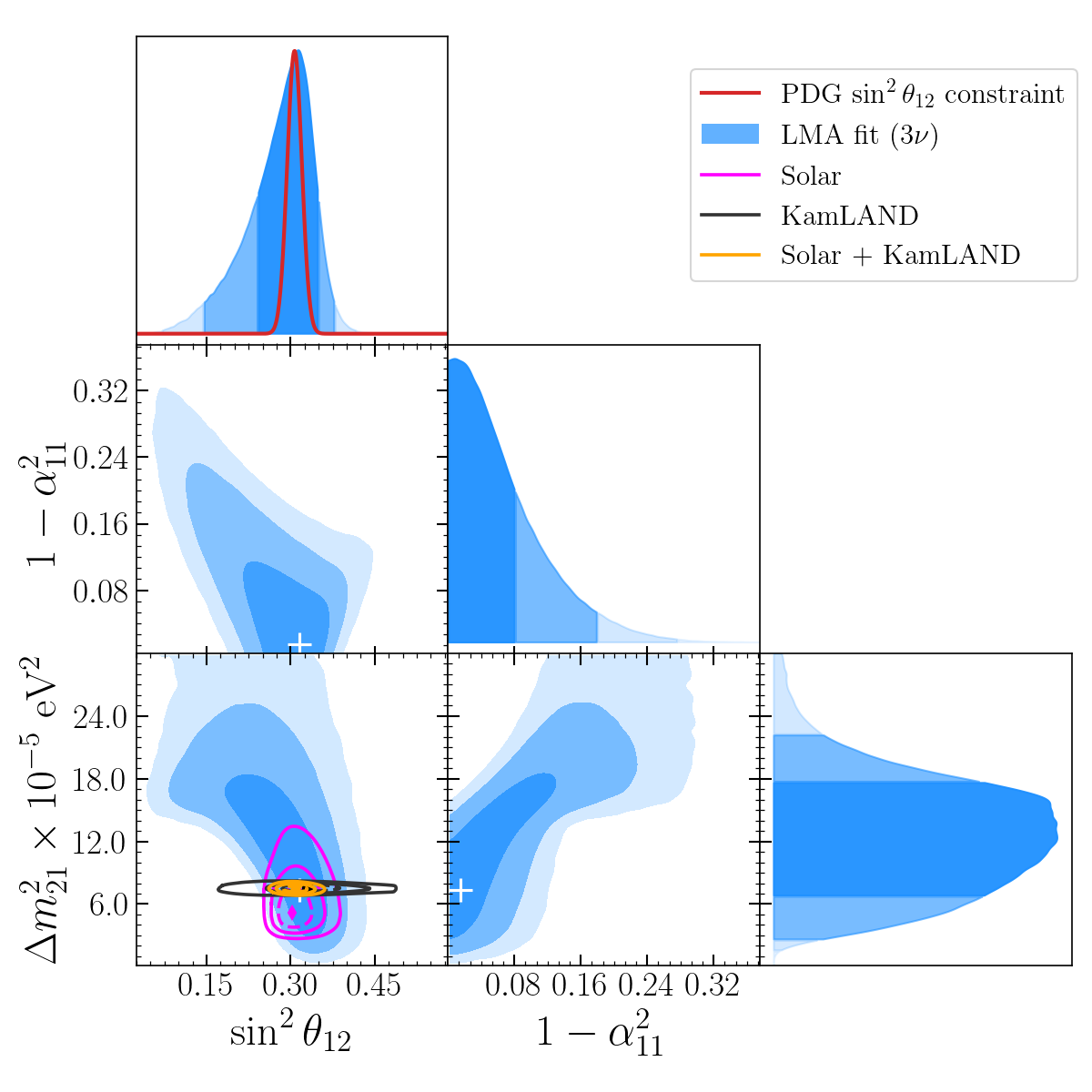}
\includegraphics[width=.47\textwidth]{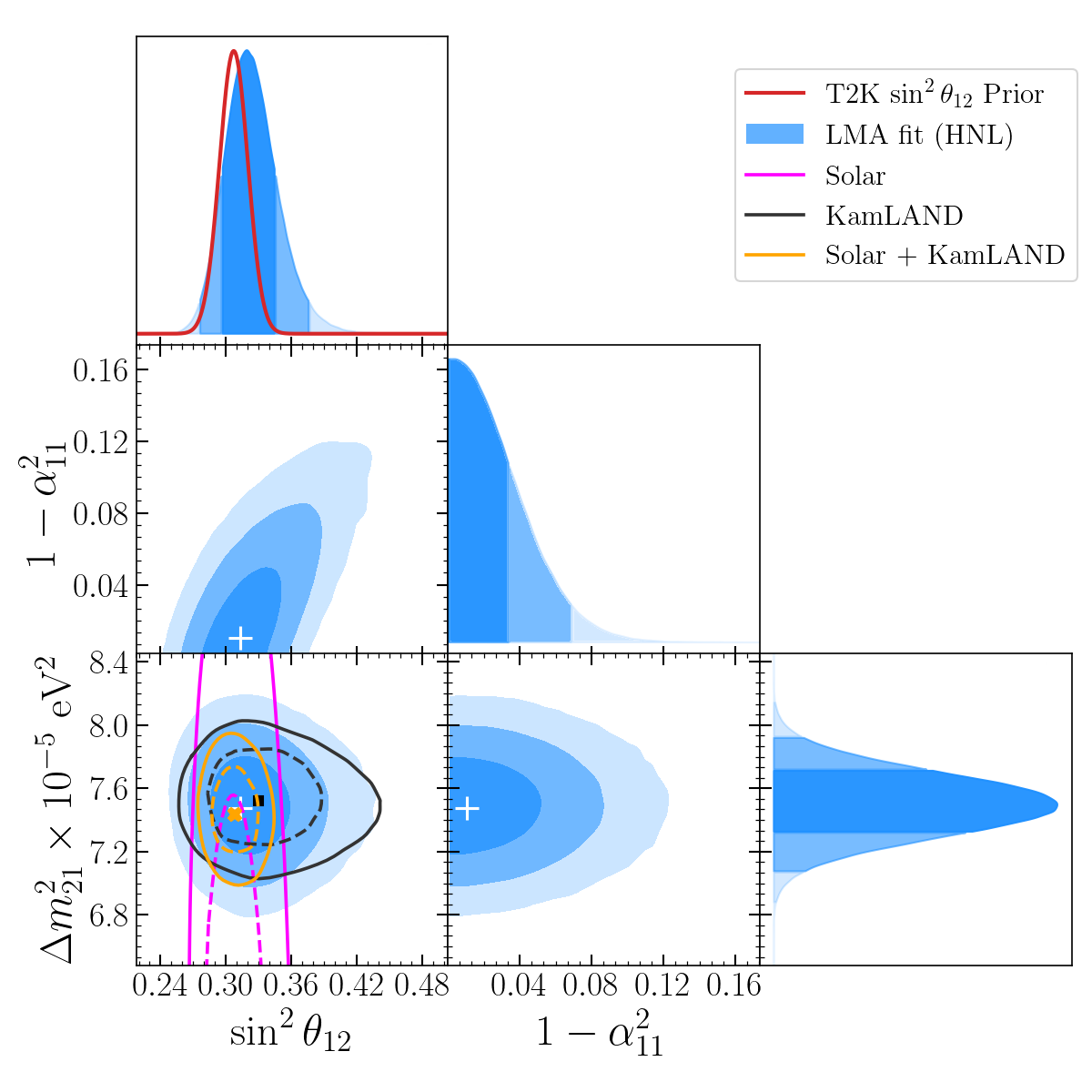}
\caption{\label{fig:HNL_fits} $1\sigma$, $2\sigma$ and $3\sigma$ contours of the posterior distributions for our Bayesian fit of the non-unitary LMA solution with a uniform prior in all parameters (left) and a Gaussian prior on $ \Delta m_{12}^2 $ (right); with $1\sigma$, $2\sigma$ (right) and $3\sigma$ (left) contours for KamLAND's result (black) and the combined result from solar experiments (magenta). The red line corresponds to the $\sin^2\theta_{12}$ constraint recommended by the Particle Data Group and the best-fit point of our analysis is shown as a white cross.}
\end{figure*}
\subsection{Non-unitary adiabatic approximation}
To proceed with an analysis we need to confirm that, at the energy and matter density scales in the sun, non-unitary neutrino evolution remains adiabatic.

The resonance takes place at $\mathcal{H}_{11} \approx \mathcal{H}_{22}$. Using the non-unitary Hamiltonian shifts the resonance from the usual condition $Ev_{cc}  = \Delta m^2_{21} \cos2\theta_{12}$ to
\begin{align}
    \frac{\alpha_{11}^4}{\alpha_{11}^2+\alpha_{22}^2}&2Ev_{cc}-(\alpha_{11}^2-\alpha_{22}^2)2Ev_{nc}= \nonumber \\ &= \Delta m_{21}^2(\cos2\theta_{12}+\mathcal{O}(|\alpha_{21}|)\sin2\theta_{12} ). \label{eqn:res_condition}
\end{align}
Notice that, as one should expect, the usual relation is recovered in the unitary limit. Then, even assuming non-unitarity on the scale of the reactor angle $\theta_{13}$, and taking (inside the Sun) $v_{nc} \approx v_{cc}/4$, the resonant density increases by no more than $\sim 20\%$ --only a few percent closer to the Sun's core~\cite{Bahcall2004}.

The adiabatic condition requires the rate of change of the effective mixing angle $\theta_{m}$ to be much smaller than the instantaneous mass splitting at resonance $\Delta H_{res}$. Due to the scale of neutrino masses, we assume a decoupling of the $1-2$ and $1-3$ resonance. Then, projecting the solar sector into a $2\times2$ space yields
\begin{align}
    \tan 2\theta_{m} \simeq \frac{2|\mathcal{H}_{12}|}{\mathcal{H}_{22} - \mathcal{H}_{11}}
    \quad\text{and}\quad
    \Delta H_{res} \simeq 2|\mathcal{H}_{12}|.
\end{align}
Thus, adiabaticity holds if
\begin{equation}
    \frac{\big|\mathcal{H}_{22} - \mathcal{H}_{11}\big|'_{res}}{8|\mathcal{H}_{12}|^2} \ll 1
\end{equation}
which, assuming unitarity, recovers the familiar
\begin{equation}
    \gamma_{\odot} \equiv \frac{E\cos(2\theta_{12})}{\Delta m_{21}^2 \sin^2(2\theta_{12})h_e} \ll 1
\end{equation}
for $h_{e}$ the coefficient of the exponential electron density profile in the sun. Substituting for $8^B$ neutrinos one finds $\gamma_\odot \sim 10^{-4}$, indicating a strongly adiabatic transition.

Setting $h_n \approx h_e$, the non-unitary expressions \ref{eqn:vacuum_hamiltonian_entries} and \ref{eqn:matter_hamiltonian_entries} yield
\begin{align}
    \hat{\gamma}_{\odot} \equiv \frac{(\alpha_{11}^2+\alpha_{22}^2)E\cos(2\theta_{12}) + \mathcal{O}(|\alpha_{21}|)}{2h_e\alpha_{11}^2\alpha_{22}^2\Delta m^2_{21}\sin^2\theta_{12}} \ll 1.
\end{align}

Once again assuming a deviation from unitarity as large as the reactor angle, we find that $\hat{\gamma}_{\odot}$ is only about $4\%$ larger than $\gamma_\odot$. The problem is still very strongly adiabatic.

\subsection{Non-unitary survival probability}
In order to find an approximation analogous to that of the unitary formalism, we want to write $\mathcal{H}_{prop}$ at some energy and matter density as $\hat{N}M\hat{N}^\dagger$, where $\hat{N}$ is an effective mixing matrix of the form $A\hat{U}$ with $A(\alpha_{ij})$ the same lower-triangular matrix and $\hat{U}(\hat{\theta}_{ij})$ still unitary. In this case, the expression for $\cos(2\hat{\theta}_{12})$ in equation \ref{eqn:LMA_3nu} is still a valid approximation, but the ratio $\beta$ will reflect the new matter effect and the non-unitary contribution to the vacuum Hamiltonian. Then, comparing expressions \ref{eqn:vacuum_hamiltonian_entries} and \ref{eqn:matter_hamiltonian_entries}, the new matter-to-vacuum ratio is
\begin{align}
    \beta_{NU} = \frac{2E}{\Delta m_{21}^2}\Big( \alpha_{11}^2c_{13}^2v_{cc}+(\alpha_{22}^2-c_{13}^2\alpha_{11}^2)v_{nc} + \mathcal{O}(|\alpha_{21}|)v_{nc} \Big)
    \label{eqn:beta_4nu}
\end{align}

Then, we follow the original 3-flavour eigenvalue crossing derivation in~\cite{Shi:1991zw}, where the survival probability averaged over the oscillation period can be written as 
\begin{align}
    \langle P_{ee} \rangle =& |a_{1}N_{e1}(L_0)N_{e1}(L_f)|^2 + |a_2N_{e2}(L_0)N_{e1}(L_f)|^2 \nonumber \\
                           +& |a_1N_{e2}(L_0)N_{e2}(L_f)|^2 + |a_2N_{e1}(L_0)N_{e2}(L_f)|^2 \nonumber \\
                           +& |a_3N_{e3}(L_0)N_{e3}(L_f)|^2
\end{align}
for $N_{\alpha i}(L_0),$ $N_{\alpha i}(L_f)$ elements of the effective mixing matrices at production and detection respectively, and $a_{i}$ the resonance couplings. Since the $\nu_1/\nu_2$ resonance appears at much lower energies than the $\nu_2/\nu_3$ resonance, we assume they are fully decoupled so that $|a_1|^2=\cos^2\theta_{13}P_{jump}$, $|a_2|^2 = 1 - \cos^2\theta_{13}P_{jump}$ and $|a_3|^2=1$; where $P_{jump}$ is probability for a neutrino to jump between mass states at the eigenvalue crossing. Using $\hat{N}$ and $N$ for the effective mixing matrices in matter and vacuum, the survival probability becomes
\begin{align}
    \langle P_{ee} \rangle =& |a_1\hat{N}_{e1}N_{e1}|^2 + |a_2\hat{N}_{e2}N_{e1}|^2 \nonumber \\
                           +& |a_1\hat{N}_{e2}N_{e2}|^2 + |a_2\hat{N}_{e1}N_{e2}|^2 \nonumber \\
                           +& |a_3\hat{N}_{e3}N_{e3}|^2.
\label{eqn:LMA intermediate step}
\end{align}
Once again exploiting the energy scale of the $\nu_1/\nu_2$ resonance, $\nu_{3}$ is strongly decoupled~\cite{Shi:1991zw} so that $\hat{\theta}_{13}\approx\theta_{13}$ and $\hat{\theta}_{23}\approx\theta_{23}$ and equation \ref{eqn:LMA intermediate step} simplifies to
\begin{align}
    \langle P_{ee} \rangle &= \alpha_{11}^4\Big( c_{13}^4\Big[ \frac{1}{2} + \Big(\frac{1}{2} - P_{jump}\cos^2\theta_{13} \Big)\nonumber \\ &\;\;\;\;\times \cos2\theta_{12}\cos2\hat{\theta}_{12} \Big]  + s_{13}^4 \Big).
    \label{eqn:final_step_LMA_derivation}
\end{align}

For the Sun's density profile, the transition is adiabatic: $P_{jump}\approx 0$~\cite{Shi:1991zw}\cite{Masses}, and so we end up with 
\begin{equation}
\langle P_{ee} \rangle = \alpha_{11}^4\Big(\frac{1}{2}\cos^4\theta_{13}[1 + \cos2\hat\theta_{12}\cos2\theta_{12}] + \sin^4\theta_{13}\Big)
    \label{eqn:LMA_4nu}
\end{equation}
which resembles the usual LMA solution, with an additional $\alpha_{11}^4$ factor. Here $\cos2\hat\theta_{12}$ must be calculated using the $\beta_{NU}$ from equation \ref{eqn:beta_4nu}.

Finally, notice that the dominant term in equation~\ref{eqn:beta_4nu} is $\alpha_{11}^2c_{13}^2v_{cc}$, as the neutral current contribution $\alpha_{22}^2-c_{13}^2\alpha_{11}^2$ is of the scale of the off-diagonal elements of $A$, and further suppressed by $v_{cc}$ being around a factor of $4$ smaller than $v_{nc}$ in the Sun. This implies that the non-unitary survival probability for $^8$B neutrinos is only very weakly dependent on any new parameters other than $\alpha_{11}$. We can approximate the behaviour of the sub-leading neutral current contribution while only incurring a $\mathcal{O}(|\alpha_{21}|)$ error as
\begin{equation}
    \beta \approx \frac{\sqrt{2}G_f E}{\Delta m_{21}^2}\Big( 2\alpha_{11}^2c_{13}^2 N_{e} + (1-c_{13}^2\alpha_{11}^2)N_n\Big)
    \label{eqn:beta_4nu_simplified}.
\end{equation}

Figure \ref{fig:LMAtest} shows the unitary and non-unitary LMA approximations against numerical solutions for increasingly large non-unitary mixing at the energy and matter density ranges of solar neutrinos. Even for large unitarity violation ($1-\alpha_{ii}<0.1$), the error in our approximation remains sub-percent.

\section{Non-unitary solar oscillation fits \label{sec:nonU-solar}}
The MSW approximation in the presence of HNLs was fitted to solar and reactor data. Taking $N_e$ and $N_n$ as given in the neutrino production regions of the sun by the BS05 SSM, we performed a Bayesian fit to the same solar data used for the unitary fit in section \ref{sec:solarConstraint}. We note that KamLAND's experimental setup is such that matter effects play no role in its oscillation measurement, so any non-unitary effects in its $\Delta m_{21}^2$ constraint come from non-unitary effects in the vacuum Hamiltonian. Since HNLs only appear as a normalisation factor in vacuum oscillations, their presence does not affect the period of the oscillations; it is simple to see that the (already angle-uncorrelated) $\Delta m_{21}^2$ measurement at KamLAND is not affected under this relaxation of the unitarity condition. Taking this into account, we use the KamLAND $\Delta m_{21}^2$ constraint as a prior for our non-unitary analysis.

Figure \ref{fig:HNL_fits} shows the result of the fits with and without KamLAND's $\Delta m_{21}^2$ constraint. The non unitary parameter is written as $1 - \alpha_{11}^2$ because in a 3+1 scenario this quantity corresponds to the $\sin^2\theta_{14}$ angle. Without the KamLAND constraint, the non-unitary parameter is strongly correlated with the solar mass splitting; this is an expected feature because a larger $\Delta m_{12}^2$ will move the transition point towards higher energies, thus increasing the survival probability for $^8$B neutrinos---an excess that can be balanced by $\alpha_{11}<1$ acting as a normalisation factor decreasing the expected flux.

\begin{table}[h!]
\centering
\resizebox{0.48\textwidth}{!}{
\renewcommand{\arraystretch}{1.2}
\begin{tabular}{lcc}
Data set      & 90\% C.L & 99\% C.L \\ \hhline{===}
NOMAD + NuTeV \cite{Escrihuela2015} & $<0.031$   & $<0.056$   \\ \hline
\begin{tabular}[c]{@{}l@{}}Global seesaw fit \\(+LFT) \cite{Fernandez_Martinez_2016}\cite{Escrihuela2017}\end{tabular} & $<2.6\times 10^{-3}$ & $<3.78 \times 10^{-3}$ \\ \hline
\begin{tabular}[c]{@{}l@{}}Global oscillation fit\\ (SBL + LBL + reactor) \cite{Parke_2016}\end{tabular}    & $<0.02$              & $<0.05$                \\ \hline
\textbf{This work}    & $<0.028$   & $<0.046$   \\ \hline
\end{tabular}
}

\caption{Curent constraints for $1-\alpha_{11}$ at the $90\%$ and $99\%$ confidence (credible) levels. The Global seesaw fit (second row) uses lepton flavour violation (LFT) data and is therefore not an oscillation constraint. The global oscillation fit uses a combination of short-baseline (SBL), long-baseline, and reactor data. The top-row measurement considers SBL accelerator data only.}\label{tab:results}
\end{table}

The results are consistent with no HNL mixing, and using the KamLAND mass difference we achieve constraints comparable to the current strongest limits (coming from joint-fits of reactor and short-baseline data \cite{Forero2021}\cite{Escrihuela2017}\cite{Blennow_2017}), constraining $(1-\alpha_{11}) < 0.046$ at $99\%$ credibility. Table \ref{tab:results} compares these results with current constraints coming from reactor, accelerator and lepton universality data; the solar limit can compete with other oscillation-only constraints and may contribute to strengthening the global limit. Unsurprisingly, introducing non-unitary parameters comes at the cost of reducing our sensitivity to $\theta_{12}$; moreover, the posteriors for $\alpha_{11}$ and $\sin^2\theta_{12}$ are highly correlated---this is further indication that the usual solar constraint used by long-baseline experiments is not adequate for HNL studies.

\begin{figure}
\centering
    \includegraphics[width=0.47\textwidth]{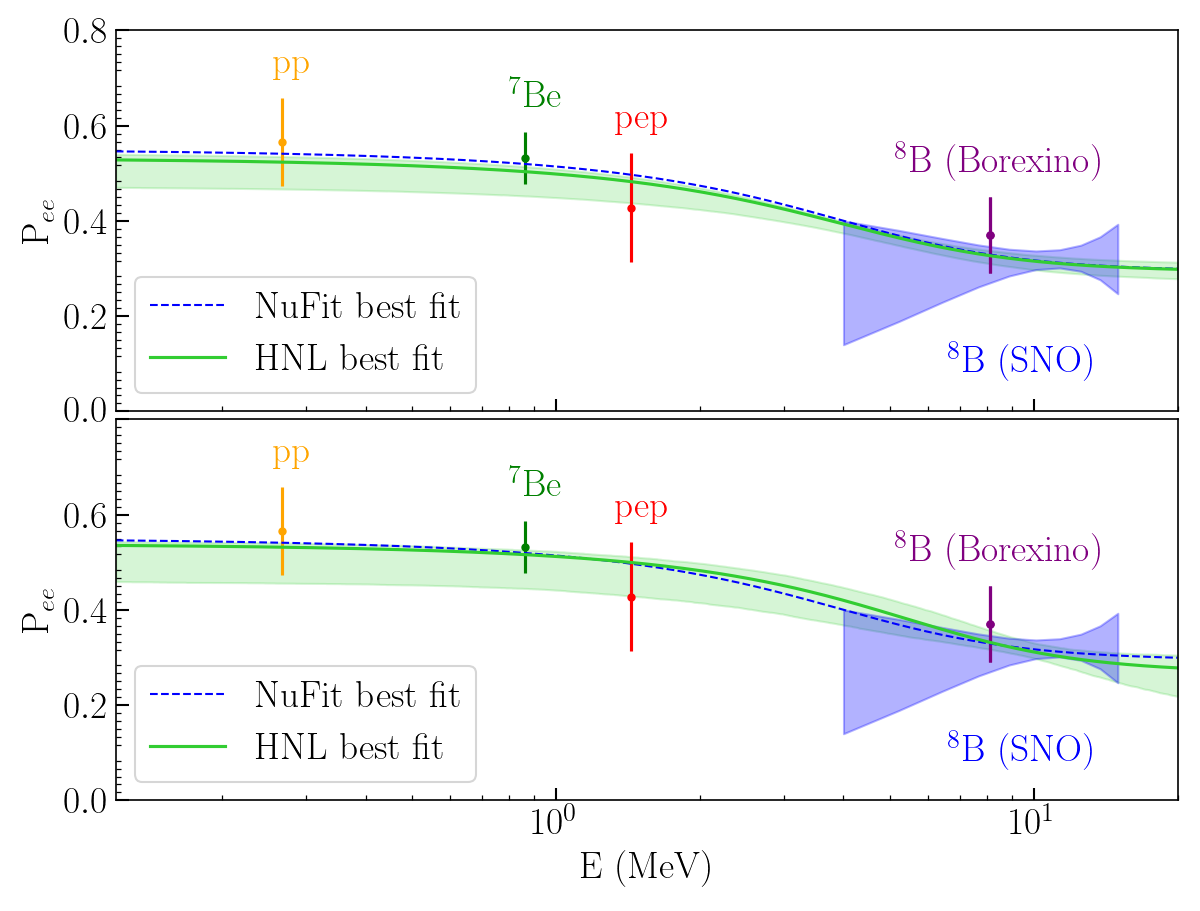}
    \caption{Solar survival probabilities from Borexino and SNO. The highest posterior density point and 1$\sigma$ contours for our non-unitary Bayesian fit with (above) and without (below) a Gaussian prior on $\Delta m_{21}^2$ are shown in green. The LMA solution for NuFit's global best-fit values is shown in blue for reference. Note that the best-fit line is not centred around the $1\sigma$ contours because the posteriors are non-Gaussian.}
    \label{fig:HNL_survival}
\end{figure}

Mirroring the presentation of the unitary fits, figure \ref{fig:HNL_survival} shows the HNL fits in $\nu_e$ survival probability space. As expected, the contours are similar to those in figure \ref{fig:3nu_survival}; differing most at the MSW transition in the $\approx 3$ MeV range. Enhanced measurements of $^8$B and pep neutrinos in this energy range may allow for stronger constraints on non-unitary $\nu_e$ mixing.

\section{Conclusion \label{sec:conclusion}}
In this work we have derived an approximation for solar survival probabilities under the HNL formalism discussed in \cite{Escrihuela2015}. The approximation introduces one new parameter and is accurate for small deviations from unitarity. We have discussed the importance of an external constraint on $\theta_{12}$ for LBL oscillation analysis and have reproduced current constraints using solar and reactor data. We have used the newly derived non-unitary approximation to constrain the $\alpha_{11}$ parameter and provided a correlated $\theta_{12}$ constraint, which is necessary for non-unitary LBL analyses.

The result has been a weakening of the current $\theta_{12}$ measurement but a competitive constraint on $\alpha_{11}$ when compared against short-baseline and global oscillation fits. Due to the role of the $\theta_{12}$ constraint on $\delta_{CP}$ measurements in LBL experiments, we can expect a decline in sensitivity to CP-violation when performing HNL fits using this work as the external solar constraint.

The fits presented in this work used publicly available data and lacked the energy-dependent information necessary to reject the low-$\theta_{12}/$high-$\Delta m_{21}^2$ region of the three flavour LMA phase space without the help of KamLAND's $\Delta m_{21}^2$ constraint, but we have shown that solar data can produce competitive constraints on the mixing between $\nu_e$ and the HNL sector and have provided a tentative constraint for unitarity violation in LBL oscillations. The non-unitary solar constraint presents strong correlations between the solar mixing angle and the non-unitary parameter $\alpha_{11}$, which reinforces the need for a special solar constraint when performing HNL searches using LBL oscillation data.

Finally, we note that the presence of non-unitarity relaxes the solar-KamLAND $\Delta m_{21}^2$ tension by allowing a larger mass splitting in solar measurements. Indeed, the highest posterior density of our solar-data-only HNL fit agrees with the KamLAND $\Delta m_{21}^2$ best-fit point, which should be blind to HNL effects. 

The solar sector has the power to set strong limits on non-unitary neutrino mixing that have wider implications in explaining nuclear anomalies \cite{Denton_2023_atomki} and is vital for accessing the off-diagonal entries of the non-unitary matrix in LBL experiments. We look forward to in-depth solar HNL analyses, particularly with the inclusion of data from the Super-Kamiokande experiment \cite{SKsolar}, anticipating leading constraints on the magnitude of $\nu_e$ non-unitary mixing and a possible resolution of the KamLAND $\Delta m_{21}^2$ tension.

\section{Acknowledgements}
I would like to thank Lukas Berns and Mariam T\'ortola for providing helpful conversations and advice on the nuances of the non-unitary oscillation formalism, Jeanne Wilson and Daniel Cookman for insight into solar neutrino physics at SNO, and Asher Kaboth, Nikolaos Kouvatsos, Teppei Katori and Francesca di Lodovico for useful comments on drafts for this work.

\nocite{*}

\bibliography{apssamp}

\end{document}